\documentstyle[12pt]{article}
\catcode`\@=11
\long\def\@makefntext#1{ %\parindent 1em
\protect\noindent \hbox to 3.2pt {\hskip-.9pt  
$^{{\ninerm\@thefnmark}}$\hfil}#1\hfill} %can be used 

 \def\@makefnmark{\hbox to 0pt{$^{\@thefnmark}$\hss}}  %original 
	
\def\ps@myheadings{\let\@mkboth\@gobbletwo
\def\@oddhead{\hbox{} %\sl
\rightmark\hfil\ninerm\thepage}   
\def\@oddfoot{}\def\@evenhead{\ninerm\thepage\hfil %\sl
\leftmark\hbox{}}\def\@evenfoot{}
\def\sectionmark##1{}\def\subsectionmark##1{}}

\newcounter{sectionc}\newcounter{subsectionc}\newcounter{subsubsectionc}
\renewcommand{\section}[1] {\vspace{0.6cm}\addtocounter{sectionc}{1} 
\setcounter{subsectionc}{0}\setcounter{subsubsectionc}{0}\noindent 
	{\bf\thesectionc. #1}\par\vspace{0.4cm}}
\renewcommand{\subsection}[1] {\vspace{0.6cm}\addtocounter{subsectionc}{1} 
	\setcounter{subsubsectionc}{0}\noindent 
	{\it\thesectionc.\thesubsectionc. #1}\par\vspace{0.4cm}}
\renewcommand{\subsubsection}[1] {\vspace{0.6cm}\addtocounter{subsubsectionc}{1}
	\noindent {\rm\thesectionc.\thesubsectionc.\thesubsubsectionc. 
	#1}\par\vspace{0.4cm}}

\newcounter{appendixc}
\newcounter{subappendixc}[appendixc]
\newcounter{subsubappendixc}[subappendixc]

\renewcommand{\appendix}[1] {\vspace{0.6cm}
        \refstepcounter{appendixc}
        \setcounter{figure}{0}
        \setcounter{table}{0}
        \setcounter{equation}{0}
        \renewcommand{\thefigure}{\Alph{appendixc}.\arabic{figure}}
        \renewcommand{\thetable}{\Alph{appendixc}.\arabic{table}}
        \renewcommand{\theappendixc}{\Alph{appendixc}}
        \renewcommand{\theequation}{\Alph{appendixc}.\arabic{equation}}
        \noindent{\bf Appendix \theappendixc #1}\par\vspace{0.4cm}}

\def\abstracts#1{{
	\centering{\begin{minipage}{30pc}\tenrm\baselineskip=12pt\noindent
	\centerline{\tenrm ABSTRACT}\vspace{0.3cm}
	\parindent=0pt #1
	\end{minipage} }\par}} 

\renewenvironment{thebibliography}[1]
	{\begin{list}{\arabic{enumi}.}
	{\usecounter{enumi}\setlength{\parsep}{0pt}
\setlength{\leftmargin 1.25cm}{\rightmargin 0pt}
	 \setlength{\itemsep}{0pt} \settowidth
	{\labelwidth}{#1.}\sloppy}}{\end{list}}

\newcounter{itemlistc}
\newcounter{romanlistc}
\newcounter{alphlistc}
\newcounter{arabiclistc}

\newcommand{\fcaption}[1]{
        \refstepcounter{figure}
        \setbox\@tempboxa = \hbox{\tenrm Fig.~\thefigure. #1}
        \ifdim \wd\@tempboxa > 6in
           {\begin{center}
        \parbox{6in}{\tenrm\baselineskip=12pt Fig.~\thefigure. #1 }
            \end{center}}
        \else
             {\begin{center}
             {\tenrm Fig.~\thefigure. #1}
              \end{center}}
        \fi}

\newcommand{\tcaption}[1]{
        \refstepcounter{table}
        \setbox\@tempboxa = \hbox{\tenrm Table~\thetable. #1}
        \ifdim \wd\@tempboxa > 6in
           {\begin{center}
        \parbox{6in}{\tenrm\baselineskip=12pt Table~\thetable. #1 }
            \end{center}}
        \else
             {\begin{center}
             {\tenrm Table~\thetable. #1}
              \end{center}}
        \fi}

\def\@citex[#1]#2{\if@filesw\immediate\write\@auxout
	{\string\citation{#2}}\fi
\def\@citea{}\@cite{\@for\@citeb:=#2\do
	{\@citea\def\@citea{,}\@ifundefined
	{b@\@citeb}{{\bf ?}\@warning
	{Citation `\@citeb' on page \thepage \space undefined}}
	{\csname b@\@citeb\endcsname}}}{#1}}

\newif\if@cghi
\def\cite{\@cghitrue\@ifnextchar [{\@tempswatrue
	\@citex}{\@tempswafalse\@citex[]}}
\def\citelow{\@cghifalse\@ifnextchar [{\@tempswatrue
	\@citex}{\@tempswafalse\@citex[]}}
\def\@cite#1#2{{$\null^{#1}$\if@tempswa\typeout
	{IJCGA warning: optional citation argument 
	ignored: `#2'} \fi}}

\def\fnt#1#2{\footnotetext{\kern-.3em
	{$^{\mbox{\sevenrm #1}}$}{#2}}}

 1
\font\twelverm=cmr10 scaled\magstep 1
 1

\font\tenbf=cmbx10
\font\tenrm=cmr10
\font\tenit=cmti10

\font\ninerm=cmr9

\def\hi{\vphantom{\int\limits_0^0}}
\def\lo{\vphantom{\int\limits}}
\begin{document}
\centerline{\tenbf QUANTUM TRANSPORT THEORY}
\centerline{\ninerm Beyond Molecular Dynamics}
\vspace{0.8cm}
\centerline{\tenrm Peter A. Henning}
\baselineskip=13pt
\centerline{\tenit Theoretical Physics, 
       Gesellschaft f\"ur Schwerionenforschung GSI,
       P.O.Box 110552}
\baselineskip=12pt
\centerline{\tenit D-64220 Darmstadt, Germany}
\centerline{\tenrm \makeatletter
       E-mail: P.Henning@gsi.de \makeatother}
\vspace{0.9cm}
\abstracts{
The problem of thermalization of a quark -- gluon plasma
is addressed in the framework of thermal field theory. Within
a simple approximation, the full quantum relaxation problem
is solved and compared to the Boltzmann solution. Memory
effects and a slowdown of the relaxation process are the
results, they can be partially described by using
a generalized kinetic equation without quasi-particle
approximation.}
\footnote[0]{
Proc. of the 4th International Workshop on Thermal Field Theories,
Dalian (China) 1995}

\vfil
%\vspace{0.8cm}
\twelverm   %modified by CLee 23/07/93
\baselineskip=14pt

%%%%%%%%%%%%%%%%%%%%%%%%%%%%%%%%%%%%%%%%%%%%%%%%%%%%%%%%%%%%%%%%%%%%%%%%%%%
\section{Introduction}
In the past decade the effort invested in ultrarelativistic heavy ion 
collisions (URHIC) has grown considerably \cite{QM95}. The general
hope is, that at some time in the near future one may be able to observe an
excursion of strongly interacting matter from the state 
of hadrons before the collision into the phase of a quark--gluon plasma
(QGP). Consequently, the discussion of possible signals from such
a shortlived state is quite vivid: Weakly interacting probes
like photons or lepton pairs, as well as strongly interacting
signals like those presented by quark flavors of higher mass have
been proposed. Similar to most of these
investigations is the assumption of a {\em thermalized\/} plasma
phase, followed by the calculation of the time
evolution along one or the other line of physical reasoning.

The present paper is a study of the 
{\em time scales\/} necessary for such a thermalization.
Ultimately, it is the goal to investigate a
physical scenario that one may reach in future URHIC: A sea of gluons,
initially at low temperature, is heated to a very high temperature
over a short time. In this {\em hot glue\/}, 
quark-antiquark pairs are popping up -- until at the very end a thermal
equilibrium in the sense of a degenerate plasma is reached. 
For the purpose of this conference contribution however, the
calculations will be presented on a more abstract level.
The full consideration is the subject of an extended 
paper\cite{h95neq}.

The primary motivation for this study are serious doubts that
the requirements for the applicability of standard transport theory 
( = kinetic gas theory) are fulfilled in a QGP: The thermal
scattering of constituents occurs so frequently, that
subsequent collisions overlap quantum mechanically.
This implies, that a treatment in terms of quasi-particles
is inadequate, one has to account for a nontrivial spectral function
of the system components \cite{L88,h94rep}

The use of a finite temperature field theoretical formulation
with continuous spectral function is also suggested by
the Narnhofer-Thirring theorem\cite{NRT83}, 
which states, that interacting systems at finite
temperature {\em cannot\/} be described by particles with a sharp
dispersion law. As an additional benefit, this 
approach is free of unphysical infrared singularities
occuring in standard perturbation theory.

The paper is organized as follows: In the next section a brief
introduction to the formalism necessary for non-equilibrium
quantum fields is given. In section 3 the approximate spectral
function is discussed, followed by a solution of the quantum transport equation
in section 4. In section 5 a generalized kinetic
equation is solved which stands between the usual Boltzmann equation and the
quantum transport equation of section 4.
Conclusions are drawn in the final section of the present work.

%%%%%%%%%%%%%%%%%%%%%%%%%%%%%%%%%%%%%%%%%%%%%%%%%%%%%%%%%%%%%%%%%%%%%%%%%%%
\section{Matrix-valued Schwinger-Dyson equation}
As has been pointed out by various authors,
the description of dynamical (time dependent) quantum phenomena
in a statistical ensemble necessitates a formalism with a doubled
Hilbert space\cite{D84a,RS86,LW87}.  For the present purpose the
relevant content of this formalism is, that its two-point Green
functions are 2$\times$2 matrix-valued. It is left to the reader
to chose either the conventional Schwinger-Keldysh,
or Closed-Time Path (CTP) Green
function formalism,\cite{SKF} or the technically
simpler method of Thermo Field Dynamics (TFD)\cite{Ubook}.

Within this matrix formulation, consider
the Schwinger-Dyson equation for the full quark propagator
$
           S =       S_0    +      S_0 \odot \Sigma \odot S
$.
Here $S_0$ is the free and $S$ the full two-point Green function
of the quark field, $\Sigma$ is the full self energy
and the generalized product of these is to be understood as
a matrix product (thermal and spinor indices) and an
integration (each of the matrices is a function of two space
coordinates). Throughout this paper the convention is used to write
space-time and momentum variables also as
lower indices, e.g. $\Sigma_{xy}\equiv \Sigma(x,y)$.

In the CTP formulation as well as in the $\alpha=1$ parameterization
of TFD\cite{hu92}, the matrix elements of $S$, $S_0$ and $\Sigma$ obey
\begin{equation}\label{sme}
S^{11}_{(0)}+S^{22}_{(0)}=S^{12}_{(0)}+S^{21}_{(0)}
\;\;\;\;\;\;\Sigma^{11}+\Sigma^{22}=-\Sigma^{12}-\Sigma^{21}
\;.\end{equation}
Therefore the four components of the Schwinger-Dyson equation are
not independent, the matrix equation can be simplified by a linear 
transformation which one may
conveniently express as a matrix product \cite{RS86,hu92}.
It achieves a physical interpretation only in the TFD formalism,
see ref. \cite{h94rep}. The transformation matrices ${\cal B}$ are 
\begin{equation}\label{lc}
{\cal B}(n) =
\left(\array{lr}(1 - n) &\; -n\\
                1     & 1\endarray\right)
\;,\end{equation}
depending on one parameter only. For example, the third term in the
Schwinger Dyson equation becomes
\begin{equation}\label{qptp}
  {\cal B}(n)\,\tau_3\,S_0\odot\Sigma\odot S\,({\cal B}(n))^{-1}
  = \left({\array{lr}  S_0^R\odot\Sigma^R\odot S^R & \mbox{something} \\
                       & S_0^A\odot\Sigma^A\odot S^A \endarray}\right)
\;.\end{equation}
Here, $\tau_3 = \mbox{diag}(1,-1)$,
$\Sigma^{R,A}$ are the retarded and advanced full self energy function,
and $S^{R,A}$ are the retarded and advanced full propagator 
(similarly for $S_0$)
\begin{eqnarray}\label{sra}\nonumber
\Sigma^R = \Sigma^{11}+\Sigma^{12}\;,\;\;\;\;
&\Sigma^A = \Sigma^{11}+\Sigma^{21}\\ 
S^R =  S^{11}-S^{12}\;,\;\;\;\; &S^A = S^{11}-S^{21}
\;.\end{eqnarray}
The diagonal elements of the transformed equation therefore
are {\em retarded\/} and {\em advanced\/} Schwinger-Dyson equation.
The off-diagonal element is a {\em transport equation\/}.

Now one switches to the mixed
(or Wigner) representation of functions depending on two
space-time coordinates: 
$
\tilde\Sigma_{XP} = \int\!\!d^4(x-y) \,
  \exp\left({\mathrm i} P_\mu (x-y)^\mu\right)\Sigma_{xy}
$ with $X = (x+y)/2$,
the $\tilde{}$-sign will be dropped henceforth. 
The Wigner transform of the convolution $\Sigma\odot G$  
is a nontrivial step. Formally it may be expressed as 
a gradient expansion
\begin{equation}\label{gex}
 \int\!\!d^4(x-y) \;
  \exp\left({\mathrm i} P_\mu (x-y)^\mu\right)\; \Sigma_{xz}\odot G_{zy}
 = \exp\left(-{\mathrm i}\Diamond\right)\,\tilde\Sigma_{XP} \,
  \tilde{G}_{XP}
\;.\end{equation}
$\Diamond$ is a 2nd order differential operator acting on both
functions appearing behind it,
$
\Diamond A_{XP} B_{XP}    =  
  \frac{1}{2}\left(\partial_X A_{XP} \partial_P B_{XP}-
                            \partial_P A_{XP}\partial_X B_{XP}\right)
$.
Obviously, this first-order term in the application of 
the infinite-order differential operator
$\exp(-{\mathrm i} \Diamond)$ is the Poisson bracket\cite{h94rep}.
Henceforth this operator is formally split into 
$\cos\Diamond-{\mathrm i}\sin\Diamond$. Similarly, one defines real
Dirac matrix-valued functions as real and imaginary part of propagator and
self energy:
\begin{equation}\label{split}
S^{R,A}_{XP} = G_{XP} \mp {\mathrm i} \pi {\cal A}_{XP} 
\;\;\;\;\;
\Sigma^{R,A}_{XP} =
  \mbox{Re}\Sigma_{XP} \mp {\mathrm i}\pi \Gamma_{XP}
\;.\end{equation}
${\cal A}_{XP}$ is the generalized spectral function of
the quantum field.

Now consider the equations obtained by action of Dirac
differential operators (= {\em inverse free propagators\/}) on the
matrix-transformed Schwinger-Dyson equation\cite{h94gl3}. The
diagonal components are 
\begin{eqnarray}\nonumber 
&&\mbox{Tr}\left[\left(
   P^\mu\gamma_\mu- m  \right)  {\cal A}_{XP}\right] =
  \cos\Diamond\,\mbox{Tr}\left[ 
  \mbox{Re}\Sigma_{XP}\, {\cal A}_{XP}
                         + \Gamma_{XP} \, G_{XP}\right]\\ 
\label{k8c}
&&\mbox{Tr}\left[\left(
   P^\mu\gamma_\mu- m   \right)  G_{XP}\right] =
  \mbox{Tr}\left[1\right] + 
  \cos\Diamond\, \mbox{Tr}\left[ 
  \mbox{Re}\Sigma_{XP} \, G_{XP}
                         -\pi^2\,\Gamma_{XP}\,{\cal A}_{XP}\right]
\;.\end{eqnarray}
Two important facts about these equations have to be emphasized. 
First notice that these equations do not in general 
admit a $\delta$-function solution
for the spectral function ${\cal A}_{XP}$ even in zero order of the
gradient expansion. This has led to erroneous statements in papers deriving
transport equations from the Schwinger-Dyson 
equation\cite{hm93}, because the right side of (\ref{k8c}) may not be
disregarded. In short terms, 
there is not such thing as a mass shell constraint 
in {\em quantum\/} transport theory !

Secondly, the equations do not contain odd powers of
the differential operator $\Diamond$. This implies, that when truncating 
the Schwinger-Dyson equation to first order in this differential 
operators (the usual order for the approximations leading to 
{\em kinetic\/} equations), the spectral function ${\cal A}_{XP}$ 
may still be obtained as the solution of an algebraic equation.

The off-diagonal component of the transformed Schwinger-Dyson equation
reads, after acting on it with the inverse free propagator 
\cite{h94rep,h94gl3}
\begin{equation}\label{k5}
\widehat{S}^{-1}_0   S^K_{xy}     =  \Sigma^R_{xz} \odot S^K_{zy}
        -              \Sigma^K_{xz} \odot S^A_{zy}
\;,\end{equation}
with kinetic components
$S^K  = \left( 1- n\right)\,S^{12} + n\,S^{21}$
and 
$\Sigma^K = \left( 1- n\right)\,\Sigma^{12} + 
        n\,\Sigma^{21}$.
Inserting the real functions defined before, this leads to
a differential equation, which henceforth is
labeled {\em quantum transport equation\/}\cite{h94rep,h94gl3}: 
\begin{eqnarray}\label{tpe1} \nonumber
\mbox{Tr}\left[\left(
  \partial_X^\mu\gamma_\mu + 2 \sin\Diamond\;
  \mbox{Re}\Sigma_{XP}
   + \cos\Diamond\;2\pi\Gamma_{XP}
  \right)  S^K_{XP}\right]& = \\
   2{\mathrm i} \mbox{Tr}\left[
             {\mathrm i}\sin\Diamond\;\Sigma^K_{XP} \, G_{XP}
             - \cos\Diamond\;\Sigma^K_{XP} \, 
    {\mathrm i}\pi{\cal A}_{XP}\right]&
\;.\end{eqnarray}
Note, that here even as well as odd powers of the operator
$\Diamond$ occur. The solution in zero order $\Diamond$ is
not trivial, since it leads to the diagonalization of the propagator
in equilibrium states\cite{h94rep,hu92}. 

%%%%%%%%%%%%%%%%%%%%%%%%%%%%%%%%%%%%%%%%%%%%%%%%%%%%%%%%%%%%%%%%%%%%%%%%%%%
\section{Effective fermion propagator and spectral functions}
In a thermal equilibrium state at temperature  $T$, the full
propagator of a fermionic quantum field has to obey the 
Kubo-Martin-Schwinger boundary condition\cite{KMS,LW87,hu92}:
\begin{equation}\label{kmf}
\left(1 - n_F(E)\right)S^{12}_{\mbox{\small eq}}(E,\vec{p}) +
 n_F(E) S^{21}_{\mbox{\small eq}}(E,\vec{p}) = 0
\;.\end{equation}
$n_F(E)$ is the Fermi-Dirac equilibrium distribution function at 
temperature $T$,\\ 
$n_F(E) = (\mbox{e}^{ \beta (E-\mu)}+1)^{-1}
$.
As seen above, the matrix valued propagator has only
three independent components, two of which are furthermore complex 
conjugate. One may now use the KMS condition to eliminate the off-diagonal
component of the equilibrium 
propagator in favor of $n(E)$\cite{h94rep,hu92}:
\begin{eqnarray}\nonumber
&&S^{(ab)}_{\mbox{\small eq}}(p_0,\vec{p}) =
   \int\limits_{-\infty}^\infty\!\!dE\,
        {\cal A}(E,\vec{p})\;\times \\ \label{fsk1}
&&\tau_3\, ({\cal B}(n(E)))^{-1}\;
   \left(\!{\array{ll}
         {\displaystyle \frac{1}{p_0-E+{\mathrm i}\epsilon}} & \\
    &    {\displaystyle \frac{1}{p_0-E-{\mathrm i}\epsilon}}
\endarray}\right)\;
 {\cal B}(n(E))
\;.\end{eqnarray}
Here ${\cal A}(E,\vec{p})$ is the spectral function of the
quark field, properly normalized and approaching a $\delta$-function for
vanishing interaction.

With the present paper one is addressing non-equilibrium states.
For such states one may {\em not\/} derive a spectral representation
of the propagator in general\cite{Ubook}, but one may still exploit
the fact that retarded and advanced propagator are
by definition analytical functions of the
energy parameter in the upper or lower complex energy half plane.

Hence, even for non-equilibrium states one may write in the
mixed (or Wigner) representation
\begin{equation}\label{rapf}
S^{R,A}(E,\vec{p},X)  = \mbox{Re}{G}_{XP} \mp \pi {\mathrm i} 
  {\cal A}_{XP} =
  \int\limits_{-\infty}^{\infty}\!\!dE^\prime\;
  {\cal A}(E^\prime,\vec{p},X)\;
   \frac{1}{E-E^\prime\pm{\mathrm i}\epsilon}
\;,\end{equation}
since this is nothing but the Wigner transform of
$
S^{R,A}_{xy} = \mp 2\pi{\mathrm i}\Theta\left(\pm(x_0-y_0)\right)
        {\cal A}_{xy}
$.

By inspection of eq. (\ref{k8c}) one finds, that only a self energy function
is needed for a full determination of the function ${\cal A}_{XP}$.
This self energy function is in general a functional of 
${\cal A}_{XP}$ again -- which then leads to a complicated
set of integro-differential equations for the self consistent determination
of the retarded and advanced propagator.

For the limited purpose of the present paper however, one makes
some physically motivated assumptions:
\begin{enumerate}
\item The self energy function for the
quarks is dominated by gluonic contributions. This is justified because the
quark-quark scattering cross section is much smaller than the quark-gluon
cross section.
\item The gluon background is dominated by external conditions, 
i.e., we neglect the back-reaction of quarks on the gluon distribution.
\item The external conditions determining the gluon field are changing
in a short time interval, and the system is translationally invariant in 
3-dimensional coordinate space.
\item One neglects the influence of anti-quarks in the spectral function.
This restriction is removed in the extended version of this paper,
ref. \cite{h95neq}.
\end{enumerate}
For these assumptions also exists a practical reason: They allow
a clean separation of various aspects of the quantum transport problem, 
whereas this separation is difficult (if not impossible)
when considering more realistic systems.

These assumptions lead to the following ansatz for the 
imaginary part of the self energy function:
\begin{equation}\label{ss1}
\pi\Gamma_{XP}\equiv\gamma^0\;\Gamma_t = \gamma^0\;g T(t)\,= \gamma^0\;g\,
     \left\{ {\array{lll}
     T_i & \mbox{if} & -\Lambda> t\\ 
     \displaystyle
        \frac{(t + \Lambda)\, T_f - t\, T_i}{\Lambda}
         & \mbox{if} & 0 > t > -\Lambda \\
     T_f & \mbox{if} & t > 0 \endarray} \right.
\;.\end{equation}
Within this ansatz the limit of $\Lambda\rightarrow 0$ is discussed
separately, it corresponds to instantaneous heating of the gluon
background. Furthermore one ensures causality by calculating
the real part of the self energy through a dispersion integral.
This integral is divergent, hence in principle one also needs a 
regularization procedure -- but the
effects of this divergence cancel in the equations.

For the quark spectral function, one uses the simple form
\begin{equation} 
{\cal A}(E,\vec{p},t) = \frac{\gamma^0}{\pi} 
\frac{\gamma_t}{
  \left(E - \omega_t\right)^2 + \gamma_t^2}
\;.\end{equation}
Hence, one approximates the quark spectral function by two time-dependent 
parameters $\omega_t$ and $\gamma_t$, which may be interpreted as 
effective mass and effective spectral width. 
One may argue about the validity
of this approach, in particular whether not a momentum dependent
spectral width is an absolute necessity for a realistic calculation.

However, first of all one may safely assume that the quarks 
appearing in the hot medium are slow -- hence the properties of
the quark distribution may be approximated by those of
quarks at rest. A second argument in favor of this ansatz is the
question of causality: The expectation value of the 
anti-commutator of two quark fields is nothing but the Fourier transform
of the spectral function. Hence, while for some more general 
spectral function causality may be 
violated, the above ansatz guarantees it when supplemented with a
corresponding antiparticle piece\cite{h95neq,h95comm}.

With the above spectral function the coupled system (\ref{k8c})  
reduces to {\em a single\/} nonlinear equation
for $\gamma_t$, plus the condition 
$
\omega^2_t = \omega^2_0 = \vec{p}^2 + m^2
$. 
%%%%%%%%%%%%%%%%%%%%%%%%%%%%%%%%%%%%%%%%%%%%%%%%%%%%%%%%%%%%
\begin{figure}[t]
\vspace*{75mm}
\includegraphics{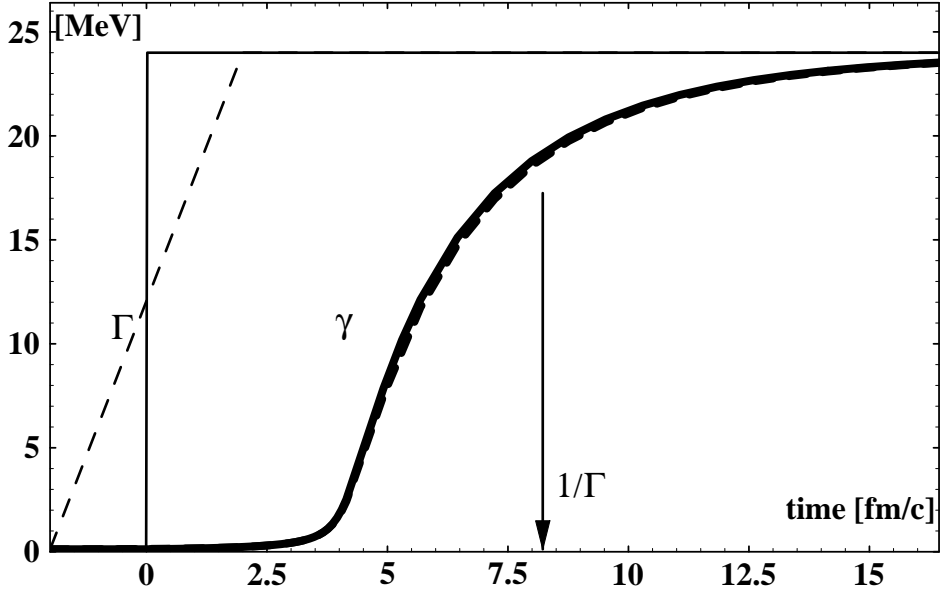}
\caption{Time dependent spectral width parameter
$\gamma_t$.\protect\newline
Parameters are $g$=0.12,  $T_i=$ 1 MeV, $T_f=$ 200 MeV,
$m=$ 10 MeV.\protect\newline
Thin lines: $\Gamma_t$ from eq. (\ref{ss1}),
thick lines: $\gamma_t$ from eqs. (\ref{k10c}), (\ref{k9c});\protect\newline
continuous lines: $\Lambda=0$, dashed lines: $\Lambda=$ 4 fm/c.
}
\vspace*{1mm}
\hrule
\end{figure}
%%%%%%%%%%%%%%%%%%%%%%%%%%%%%%%%%%%%%%%%%%%%%%%%%%%%%%%%%%%%
This latter condition is more complicated, when the anti-particle piece of 
the spectral function is taken into account\cite{h95neq}.
The energy parameter is chosen as $E=\omega_0$, which yields instead of
eq. (\ref{k8c}) as the Schwinger-Dyson equation for the retarded
(or advanced) two-point function of the quarks:
\begin{eqnarray}\nonumber
\gamma_t & = & g T_i + g (T_f - T_i)\,\Theta(t)+
                g (T_f - T_i)\,
    \left( \frac{t+\Lambda}{\Lambda} -\frac{1}{2 \gamma_t \Lambda}\right)
     \,\Theta(-t)\Theta(t+\Lambda)\\
\label{k10c}
&& \hphantom{ g T_i} -\frac{ g(T_f - T_i)}{2 \gamma_t \Lambda}\,
   \left(\Theta(t)\,{\mathrm e}^{{\displaystyle -2 \gamma_t t}}
         -\Theta(t+\Lambda)\,{\mathrm e}^{{
          \displaystyle -2 \gamma_t (t+\Lambda)}}\right)
\end{eqnarray}

In the limit $\Lambda\rightarrow 0$, this becomes even simpler:
\begin{equation}
\gamma_t = g T_i + g (T_f - T_i)\,\Theta(t)\,
   \left(1-{\mathrm e}^{{\displaystyle -2 \gamma_t t}}\right)
\label{k9c}
\end{equation}
In Fig. 1, the solution of these equations is plotted in comparison
to the time dependent imaginary part of the self energy function
from eq. (\ref{ss1}). It is obvious, that the solution of the
nonlinear equations (\ref{k10c}) resp. (\ref{k9c}) approaches
the imaginary part of the self energy function with a 
characteristic delay time. Simply using $\Gamma_t$ from eq. (\ref{ss1})
instead of $\gamma_t$ -- which would correspond to an 
{\em adiabatic\/} approximation -- therefore ignores this 
delay time. In ref.\cite{h95neq} it is discussed how this delay
time is calculated from the system parameters.

%%%%%%%%%%%%%%%%%%%%%%%%%%%%%%%%%%%%%%%%%%%%%%%%%%%%%%%%%%%%%%%%%%%%%%%%%%%
\section{Transport equation}\label{tpe}
As was stated above, the off-diagonal component of the
transformed Schwinger-Dyson equation is a transport 
equation\cite{RS86,h94rep}.
To see this more clearly, {\em define\/} the generalized
covariant distribution function $N_{XP}$ through the equation
\begin{equation}\label{nde}
\left(1-N_{XP}\right)\,S_{XP}^{12} + N_{XP}\,S_{XP}^{21} =0
\,.\end{equation}  
Note the similarity with eq. (\ref{kmf}): The above equation indeed
ensures, that in the limit of thermal equilibrium one achieves
%%%%%%%%%%%%%%%%%%%%%%%%%%%%%%%%%%%%%%%%%%%%%%%%%%%%%%%%%%%%
\begin{figure}[t]
\vspace*{75mm}
\includegraphics{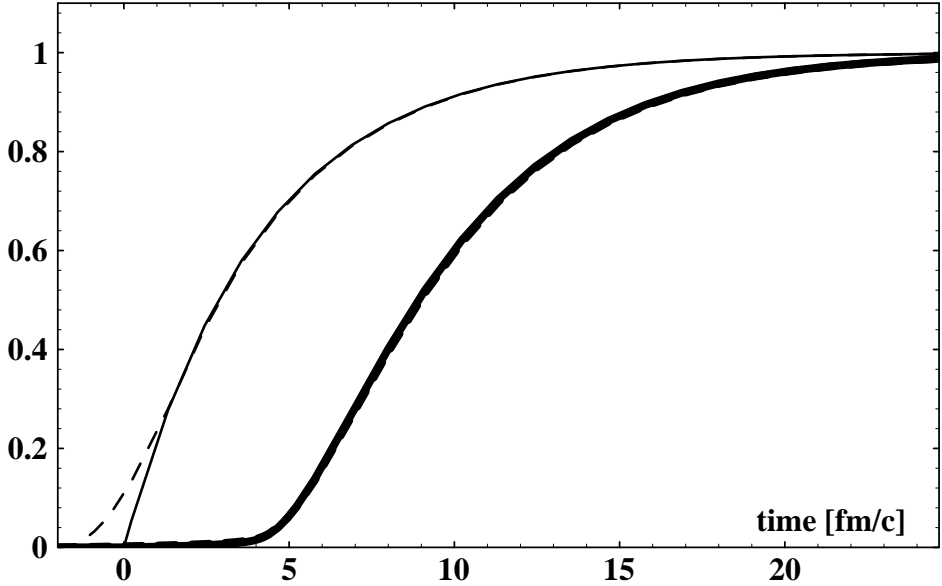}
\caption{Normalized time dependent fermionic distribution function for 
 slow quarks.\protect\newline
Parameters as in Fig. 1;
thin lines: $N^B_t/n_F(m,T_f)$ from the Boltzmann equation (\ref{tpe3}),
thick lines: $N_t/n_F(m,T_f)$ from the 
quantum transport equation (\ref{tpe2});\protect\newline
continuous lines: $\Lambda=0$, dashed lines: $\Lambda=$ 4 fm/c.
}
\vspace*{1mm}
\hrule
\end{figure}
%%%%%%%%%%%%%%%%%%%%%%%%%%%%%%%%%%%%%%%%%%%%%%%%%%%%%%%%%%%%%%%%%%%%%%%%%%%
$
\lim_{\mbox{\footnotesize equil}} N_{XP} = n_F(E)
$.
For the purpose of the present paper $N_{XP}$ is taken as a scalar 
function. The description of phenomena 
like spin diffusion requires to use a Dirac matrix valued 
$N_{XP}$\cite{hm93}.
It follows that
$
S_{XP}^K = 2\pi{\mathrm i}\,\left(N_{XP} - n\right)\,{\cal A}_{XP}
$.
From this step the mathematical interpretation of the 
generalized distribution function $N_{XP}$ is obvious: It is the
parameter which diagonalizes the the full non-equilibrium
matrix-valued propagator through the Bogoliubov matrix
${\cal B}$ from (\ref{lc})\cite{h94rep,hu92}:
\begin{equation}
{\cal B}(N_{XP})\,\tau_3\,S_{XP}\,({\cal B}(N_{XP}))^{-1}=
\left({\array{rr} G_{XP}-{\mathrm i}\pi{\cal A}_{XP} & \\
 & G_{XP}+{\mathrm i}\pi{\cal A}_{XP} \endarray}\right)
\;.\end{equation}
For the following, one furthermore defines a ``pseudo-equilibrium'' 
distribution function: The 2$\times$2 matrix structure 
of self energy function allows to diagonalize it also 
by a Bogoliubov transformation\cite{h94rep} with
a parameter $N^0_{XP}$ such that
\begin{equation}\label{psef}
\Sigma^{12}_{XP} =  2\pi{\mathrm i} N^0_{XP}\,\Gamma_{XP}\;\;\;\;\;\;\;
\Sigma^{21}_{XP} =  2\pi{\mathrm i} \left(N^0_{XP}-1\right)\,\Gamma_{XP}
\;.\end{equation}
In the present approach $N^0_{XP}$ is determined by the hot gluon gas
acting as background, hence without the back-reaction it is equal to
the equilibrium function, $N^0_{XP} \equiv n_F(E,T(t))$ with a time 
dependence due to the time dependence of the temperature.
Looking at slow quarks with $E=m$, one furthermore replaces
$N(X;m,\vec{p})$ by $N_t$ and 
neglects the energy derivative of $n_F(E,T(t))$. The resulting
quantum transport equation according to (\ref{tpe1}) then is:
\begin{equation}\label{tpe2}
\frac{d}{d t}N_t = -2 \,\gamma_t\left( N_t - n_F(m,T(t)) \right)
\;\end{equation}
with $T(t)$ as defined in eq. (\ref{ss1}).
This equation looks surprisingly similar to a kinetic equation in
relaxation time approach. However, this similarity is superficial:
The {\em kinetic\/} equation,
or Boltzmann equation, derived for this simple model system reads
\begin{equation}\label{tpe3}
\frac{d}{d t}N^B_t = -2 \,\Gamma_t\left( N^B_t - n_F(m,T(t)) \right)
\;,\end{equation}
with the imaginary part of the self energy $\Gamma_t$ from
eq. (\ref{ss1}) instead of
the spectral width parameter $\gamma_t$. That these differ
substantially in the beginning of the relaxation process has been
shown in the previous section.

Fig. 2 depicts the influence of this difference on the solution of the
transport equation. The result is that the {\em relaxation process\/}
is slowed by the inclusion of the spectral function
of the system components. Please observe, that the curves of
Fig. 2 employ the same behavior as seen in Fig.1: The
relaxation {\em rate\/} is similar in the quantum transport and the
Boltzmann equation, but the former experiences
a characteristic {\em delay time\/} with respect to the latter.
This delay time is almost doubled with respect to the delay time
occuring in the spectral width parameter $\gamma_t$, an
asymptotic calculation is carried out in ref.\cite{h95neq}.

%%%%%%%%%%%%%%%%%%%%%%%%%%%%%%%%%%%%%%%%%%%%%%%%%%%%%%%%%%%%%%%%%
\section{Gradient expansion}
One may now raise the question, whether one can produce an equation
which at least takes some of the quantum features of particles 
into account in an otherwise kinetic picture. The reason for this is,
that in a general non-equilibrium system one cannot hope to
reduce the equations (\ref{k8c}) and (\ref{tpe1}) to such simple
forms as obtained above. Even a purely numerical solution
of these equations seems to be impractical if not impossible.

Therefore, to answer the question, consider the two steps which are
between the equations (\ref{tpe2}) and (\ref{tpe3}): First of 
all a quasi-particle approximation, secondly an expansion
of the operator $\exp(-{\mathrm i}\Diamond)$ to first order,
i.e., replacing it by $1 - {\mathrm i}\Diamond$.
The first of these steps would be
in contradiction to the philosophy outlined in the introduction to this
work. The second step however  may be kept: To expand the
diagonal as well as the off-diagonal pieces of the original
matrix-valued Schwinger-Dyson equation to first order in
the operator $\Diamond$\cite{h94rep,h94gl3}.

The necessary differential equation for $N_{XP}$ has been derived in 
ref.\cite{h94gl3}, correct to first order in the gradient expansion 
it reads
\begin{eqnarray}
\nonumber
&&\mbox{Tr}\left[ {\cal A}_{XP} \left\{  \hi
\left( \lo P_\mu\gamma^\mu - m - \mbox{Re} \Sigma_{XP}
 \right), N_{XP} \right\}\right]\\
\nonumber
&&\;\;= {\mathrm i} \mbox{Tr}\left[\hi {\cal A}_{XP} \left( \lo
N_{XP} \Sigma^{21}_{XP} - \left( N_{XP}-1\right)
 \Sigma^{12}_{XP}\right) \right]\\
\nonumber
&&\;\; -{\mathrm i} 
\int\limits_{-\infty}^0\!\!d\tau \int\!\frac{dE}{2\pi}\,
\sin(\tau E)\,\mbox{Tr}\left[ \left\{  \hi
{\cal A}(X;P_0+E,\vec{P}),\right.\right.\\
\label{tpe1a}
&&\left. \left.\left(\lo
N_{XP} \Sigma^{21}(t+\tau/2,\vec{X};P) 
- \left(N_{XP}-1\right) \Sigma^{12}(t+\tau/2,\vec{X};P)\lo\right)\hi
\right\}_{N}\right]
\;.\end{eqnarray}
%%%%%%%%%%%%%%%%%%%%%%%%%%%%%%%%%%%%%%%%%%%%%%%%%%%%%%%%%%%%
\begin{figure}[t]
\vspace*{75mm}
\includegraphics{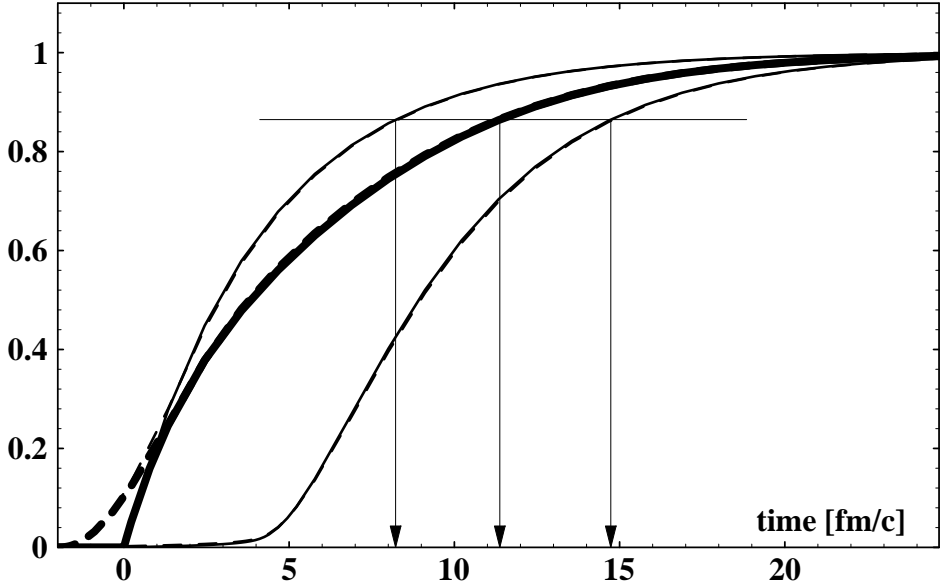}
\caption{Normalized time dependent fermionic distribution function for 
 slow quarks.\protect\newline
Parameters as in Fig. 1;
thin lines: left $N^B_t/n_F(m,T_f)$ from the Boltzmann equation (\ref{tpe3}),
right $N_t/n_F(m,T_f)$ from the quantum transport equation (\ref{tpe2});
thick lines: $N^G_t/n_F(m,T_f)$ 
from the generalized kinetic equation (\ref{tpe4}); 
continuous lines: $\Lambda=0$, dashed lines: $\Lambda=$ 4 fm/c.
}
\vspace*{1mm}
\hrule
\end{figure}
%%%%%%%%%%%%%%%%%%%%%%%%%%%%%%%%%%%%%%%%%%%%%%%%%%%%%%%%%%%%
In this equation, $\left\{\lo\cdot,\cdot\right\}$ denotes the Poisson
bracket, the index $N$ means that the derivatives 
are not acting on $N_{XP}$.  Here, as outlined before, one 
may use a spectral function which is the solution of an algebraic
equation. For the present simple model this means to replace
$\gamma_t$ by $\Gamma_t$ in the function ${\cal A}$. Note, that
the above equation is strictly causal: It involves a time integral
only over the past history of the system, and its derivation 
is based on the dispersion integral (\ref{rapf}). The problem of
unphysical singularities in the propagator therefore does not occur.

Furthermore, replacing $N_{XP}$ by the unknown function $N^G_t$ 
and inserting all the previous definitions, one obtains the nonlinear equation
\begin{eqnarray}\nonumber
\frac{d}{d t} N^G_t & = &
 - 2 \Gamma_t\,\left(N^G_t - n_F(m,T(t))\right) \\
\nonumber
&& + 2 g\left(T_f-T_i\right)\,\left[\vphantom{\int\limits_0^0}
  \Theta(t)\left(\frac{t}{\Lambda} + \frac{1}{2 \Gamma_t \Lambda}\right)
  \exp(-2 \Gamma_t t)\right.\\
\nonumber
&&-
  \Theta(t+\Lambda)\left(\frac{t+\Lambda}{\Lambda} + 
  \frac{1}{2 \Gamma_t \Lambda}\right)\exp(-2 \Gamma_t (t+\Lambda))\\
&&+\left.\vphantom{\int\limits_0^0}
  \Theta(-t)\Theta(t+\Lambda)\,\frac{1}{2\Gamma_t\Lambda}\right]\,
  \left(N^G_t - \frac{n_F(m,T_f) T_f - n_F(m,T_i) T_i}{T_f - T_i}\right)
\label{tpe4}
\;.\end{eqnarray}
In the limit $\Lambda\rightarrow 0$ this may be simplified to 
\begin{eqnarray}\nonumber
\frac{d}{d t} N^G_t & = &
 - 2 \Gamma_t\,\left(N^G_t - n_F(m,T(t))\right) \\
\nonumber
&& + 4\,t\,\Theta(t)\, \left(g\left(T_f-T_i\right)\right)^2\,
  \exp(-2 \Gamma_t t)\,\\
&&\;\;\;\;\;
  \left(N^G_t - \frac{n_F(m,T_f) T_f - n_F(m,T_i) T_i}{T_f - T_i}\right)
\label{tpe5}
\;.\end{eqnarray}
Shown in Fig.3 is the numerical solution for $N^G_t$ in comparison
to the Boltzmann solution $N^B_t$ as well as the full quantum
transport solution $N_t$. 

%%%%%%%%%%%%%%%%%%%%%%%%%%%%%%%%%%%%%%%%%%%%%%%%%%%%%%%%%%%%%%%%%%%%%%%%%%%
\section{Discussion and Conclusion}
The comparison of the three methods to describe the relaxation problem
of a quark--gluon plasma (QGP) shows, that the full quantum transport equation
results in a {\em much \/} slower equilibration process
than the Boltzmann equation.
This result is in agreement with other attempts to solve
the quantum relaxation problem\cite{D84a,h93trans}: The quantum system
exhibits a memory, it behaves in an essentially non-Markovian way.

In particular, for the physical scenario studied here,
the system ``remembers'' that it has been equilibrated some time ago.
The relaxation {\em rate\/} then is very similar to the Boltzmann rate,
but the system follows with a characteristic delay time. This delay time
depends on the system parameters in a non-algebraic way, hence
one may be subject to surprises for physical examples.

In the present quantum transport example for the QGP,
the time to reach 1-1/e${}^2\approx$ 86 \% of the equilibrium 
quark occupation number
is almost doubled (14.7 fm/c as compared to 8.2 fm/c in the 
Boltzmann case). Thus, it may be carefully stated, that
the question of the applicability of {\em standard\/}
transport theory with quasi-particles needs further investigation:
It might turn out, that quantum effects (= memory as decribed in this
contribution) substantially hinder the thermalization of a QGP
over long time scales.

One also finds, that this result holds for 
instantaneous as well as fast ($\Lambda$ = 4 fm/c) heating of the
bosonic background. Without elaboration at this point it may be stated
that the inclusion of antiquarks into the spectral function
does not change these figures substantially; it only leads
to small oscillations of the relaxation rate around the value given
in Fig. 1.

The calculated numerical value of 14.7 fm/c 
for the thermalization time of slow quarks is certainly so large,
that the cooling of the bosonic background has to be taken into account
for realistic estimates. Thus however one runs into the principal problem
of non-equilibrium quantum field theory: The solution of time-dependent 
coupled equations for the Green's functions, hardly possible in any concrete
case. A way out of this dilemma might be offered by the generalized
kinetic equation\cite{h94rep,h94gl3} (\ref{tpe1a}), which is
related to the quantum transport equation as well as to the 
Boltzmann equation: It does not contain the convolutions over
coordinate space that are hidden in the Schwinger-Dyson equation.

However, it does contain the gradient approximation of standard
transport theory -  and thus its applicability to the system
studied here is questionable, since a step function in time
certainly involves large gradients. The present comparison is therefore 
justified only through its results:
The fact, that with the generalized transport 
 equation one does at least partially describe the
memory effects in a quantum system (the characteristic time now is
11.4 fm/c) is encouraging. Applications of this transport equation to 
more complicated systems seem to be possible, at least in cases where 
one previously has used Boltzmann-like or Vlasov-like equations which also
contain this gradient expansion to first order.

As a more general remark at the end of this paper it might be added,
that the present results certainly demonstrate the importance of
solving all three components of the matrix-valued Schwinger-Dyson
equation on the same level of approximation. Using only a trivial
approximation to the diagonal equations, i.e., replacing the
spectral functions of the model by some ``mass-shell constraint'',
is not justified for strongly interacting hot systems.

%%%%%%%%%%%%%%%%%%%%%%%%%%%%%%%%%%%%%%%%%%%%%%%%%%%%%%%%%%%%%%%%%%%%%%%%%%%

\end{document}